\documentclass[referee, usenatbib]{mn2e}
\usepackage{graphicx}
\title[Revisiting the formation history of the minor-axis dust lane galaxy NGC1947]{Revisiting the formation history of the minor-axis dust lane galaxy NGC1947}
\author[M. Spavone et al.]{M. Spavone$^{1,2}$\thanks{E-mail:
spavone@na.infn.it (MS); iodice@na.astro.it (EI)}, E.
Iodice$^{2}$, R. Calvi$^{1}$, D. Bettoni$^{3}$, G. Galletta$^{4}$,
G. Longo$^{1,2}$,\and P. Mazzei$^{3}$, G. Minervini$^{1}$\\
$^{1}$Dipartimento di Scienze Fisiche, Universit\'a Federico II,
via Cinthia 6, I-80126 Napoli, Italy\\
$^{2}$INAF-Astronomical Observatory of Naples, via Moiariello 16,
I-80131 Napoli, Italy\\
$^{3}$ INAF-Astronomical Observatory of Padova, Vicolo
dell'Osservatorio 5, 35122 Padova, Italy\\
$^{4}$ Dipartimento di Astronomia, Universit\'a di Padova, Vicolo
dell-Osservatorio 2, 35122 Padova, Italy}
\begin{document}
\date{Accepted 2008 November 10 Received 2008 November 05; in original form 2008 July 29}

\pagerange{\pageref{firstpage}--\pageref{lastpage}} \pubyear{2008}

\maketitle

\label{firstpage}

\begin{abstract}
In this paper we present a detailed study of the peculiar early-type
galaxy NGC1947. The main goal of this work is to constrain the
dynamical status and the formation history of NGC1947 by comparing the
observed properties with the predictions derived from different galaxy
formation scenarios. To this aim, we derived the photometric and
kinematical properties of NGC1947. Due to the presence of an extended
dust-lane, which crosses the galaxy center along the photometric minor
axis, we used near-infrared images (J and K bands) to derive an
accurate analysis of the stellar light distribution. Optical images
(in the V and R bands) are used to derive the color profiles and color
maps to study the structure of the dust-lane. The observed kinematics
confirm the presence of two components with decoupled angular
momentum: gas and dust rotate along the minor axis, while the rotation
velocities of the stars are observed along the major axis. The complex
structure observed in NGC1947 support the hypothesis that some kind of
interactions happened in the evolution of this object. We analyzed two
alternatives: a merging process and an accretion event. We discussed
how the observed properties strongly suggest that the decoupled ring
of gas and dust have been accreted from outside.
\end{abstract}


\section{Introduction}\label{intro}

One of the major open issues in modern cosmology is to understand
how galaxies formed and evolved. It is likely that the formation
of galaxies is dominated by two processes: the assembly of
luminous and dark matter through accretion and mergers, and the
conversion of baryonic and non-baryonic matter into stars. The
Cold Dark Matter (CDM) scenario for galaxy formation is based on
the hierarchical mass assembly \citep{Cole00} which predicts that
the observed galaxies and their dark halo were formed through a
repeated merging process of smaller systems. This is the reason
why the study of galaxy interactions and merging has received an
ever increasing attention in recent years, both on the
observational and theoretical sides.

It has been shown that gravitational interactions and mergers
affect the morphology and dynamics of galaxies from the Local
Group to the high-redshift universe (e.g. \citealt{Cons03};
\citealt{Bundy04}; \citealt{Lin04}; \citealt{Kart07};
\citealt{DeP07}; \citealt{Lotz08}). These observational results
support the scenario in which the merging of two disk galaxies
produces spheroidal merger remnants with physical characteristics,
such as density profiles, mean velocity dispersion and surface
brightness, quite similar to those observed for E and S0s
(\citealt{Too72}; \citealt{Bar92}; \citealt{Bek98a};
\citealt{Bour05}). However, other types of interactions, such as
smooth accretion and stripping seem to be equally important in the
growth of galaxies and the relative weights of the various
processes depends on the environment, which drive many morphological
features observed in galaxies, such as bars, shells (see
\citealt{Mal80}) and highly inclined or even counter-rotating
rings/disks (see \citealt{Gal96} for a review). Concerning the
latter morphology, different degrees of evolution of
these peculiar rings/disks are seen: some cases the decoupled
matter is composed by stars and gas, and regions of recent star
formation are also observed (\citealt{Iod02a}, \citealt{Iod02b};
\citealt{Res05}); other cases show rings with a particularly
smooth structure \citep{Iod04}, or formed only by gas
\citep{Bet01}. In general, these peculiar galaxies have very large
amount of gas (\citealt{Ric94} and \citealt{Gal97}) and dust which
lies in the same plane of the decoupled ring/disk (e.g.
\citealt{Arna97}; \citealt{Bet01}; \citealt{VanD02};
\citealt{Cox06}). The decoupling of the angular momentum between
the two components cannot be explained by the collapse of a single
protogalactic cloud, thus a ``second event'' must have happened in
the formation history of these systems. Given all the above, such
peculiar galaxies turn out to be the ideal laboratory where to
derive detailed information about many of the physical processes
controlling galaxy interactions and merging.

Several theoretical works based on numerical simulations, have
tested two plausible scenarios for the origin of the decoupled
matter: either gas accretion from infall processes, or merging
between gas-rich galaxies (\citealt{Tha96}, \citealt{Tha98};
\citealt{Res97}; \citealt{Bek98b}; \citealt{Bour03},
\citealt{Bour05}). In both cases, the decoupled material forms a
ring (\citealt{Qui91}, \citealt{Hern93}) which after a while
settles into one of the principal planes of the gravitational
potential associated with the host galaxy (\citealt{Heis82};
\citealt{Bert91}; \citealt{Bour03}). Multiple accretion events may
also account for more complicated objects, like NGC2685 where two
orthogonal rings are observed around the main galaxy body
(\citealt{Sch02}; \citealt{Bour03}).\\ The unequal-mass merging of
two disk galaxies is the other way to build a galaxy with an
highly inclined ring/disk: the morphology and kinematics of the
merger remnants depends on the initial orbital parameters and the
initial mass ratio of the two galaxies involved in the process
(e.g. \citealt{Bour03}, \citealt{Bour05}).\\ External gas can also
be accreted from cosmic web filaments with high inclined angular
momentum relative to the main disk plane (\citealt{Dav01};
\citealt{Sem05}; \citealt{Mac05}) and recent theoretical work has
argued that this might even be the most realistic way by which
galaxies get their gas \citep{Ker05}. Finally, in the standard CDM
scenario, recent high-resolution cosmological simulations of
galaxy formation have shown that angular momentum misalignment
during the hierarchical structure formation can also lead to the
formation of highly inclined ring/disk \citep{Brook08}.

\begin{figure*}
\centering
\includegraphics[width=10cm]{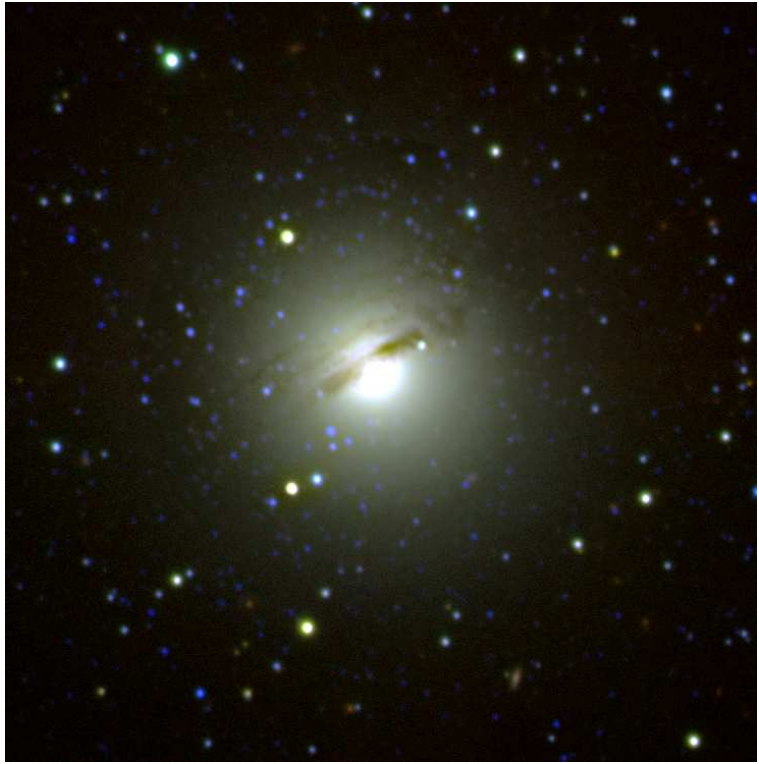}
\caption{Color composite image of NGC1947 assembled from EMMI-NTT
images in the V (blu channel), J (green channel) and K (red
channel) bands. The North is up, while the East is on the left of
the image.} \label{N1947_comp}
\end{figure*}

Although most of the formation scenarios proposed so far seem able
to account for many of the observed morphologies and kinematics
for this class of peculiar galaxies, the relative weight of the
two main processes, i.e. the merging of two galaxies or external
gas accretion, remains still uncertain. The reasons for such an
uncertainty mainly reside in the poor understanding of many
processes which are at work during gravitational interactions such
as, for instance, gas dissipation, star formation efficiency,
feedback, gas re-heating, etc., phenomena that all strongly affect
the dynamical history of galaxies. Detailed photometric and
kinematical observations of systems in different dynamical states,
both in the local universe and at higher redshift, can help better
constraining the physics of such processes. To this aim, we
present a detailed study of the peculiar early-type galaxy
NGC1947, that can be considered as a prototype of minor-axis
dust-lane galaxy (see Fig.\ref{N1947_comp}). In 1978
\citet{Bert78} made the first study of a small sample of peculiar
galaxies (including NGC1947) that all showed a dust-lane crossing
perpedicularly the central spheroidal body: the similarities in
observed properties let them to conclude that, by a morphological
point of view, these objects should be considered as a new class
of galaxies. A larger sample of dust-lane galaxies, including
objects similar to NGC1947, have been studied by \citet{Haw81},
and they found that galaxies with a minor axis dust-lane form a
dominant group among all dust-lane galaxies. The difference of the
angular momenta of the stellar, gaseous and dusty components is a
clear proof that the three components have not all the same
origin, and an accretion of new material from outside was
addressed as possible formation mechanism for such class of
objects (\citealt{Haw81}).

This work is a part of an ongoing research project aimed
to select a sample of peculiar and interacting galaxies, for
which both multiwavelength photometric and kinematical data are
available, and to compare the observed properties with the
predictions derived from different galaxy formation scenarios.

The main properties of NGC1947 are listed in Tab.\ref{General}.
The morphological classification of this galaxy is somewhat
controversial. In fact, NGC1947 was initially classified as S0
peculiar by \citet{deV76} and $S0_{3}$ peculiar by \citet{San87}
while, due to the absence of a photometrically detectable disk,
\citet{Bert92} and \citet{Oos02} classified it as an E1 crossed by
a series of dust lanes along the optical minor axis. As shown by
\citet{Bert92} the peculiarity of NGC1947 resides in the presence
of three parallel minor-axis dust lanes, which appear as
concentric rings, in addition to the central one, which is typical
of the dust-lane galaxies. The main dust lane crosses the center
along the SW direction, while the others, less pronounced and
parallel to the central one, are detected in the NE side. The
galaxy also exhibits a ring of molecular gas \citep{Sag93} and a
disk-like distribution of ionized gas centered on the nucleus and
at the same position angle of the dust lane. Moreover, the stars
in NGC1947 rotate along the galaxy's major axis, perpendicular to
the gas rotation axis \citep{Bert92}.\\ A considerable amount of
HI was detected by \citet{Oos02}: the HI gas is distributed in a
warped disk, which is aligned with the dust lane and it rotates in
the same sense; the HI mass is about $3.4 \times\ 10^{8}
M_{\bigodot}$.

This paper is structured as follows: in Section \ref{obs} we
describe the observations and data reduction procedures. The
morphology in the NIR and optical bands and the light and color
distribution of the galaxy are described in Section \ref{prof},
while in Section \ref{2D} we present a 2-dimensional model of the
light distribution. In Section \ref{kinem} we discuss the star and
gas kinematics along the two principal photometric axes of the
galaxy, and in Section \ref{sed} is presented and discussed the
spectral energy distribution derived for NGC1947.
Finally, in Section \ref{disc} we draw our conclusions.\\
Throughout this paper we shall adopt for NGC1947 a distance of
about 15 Mpc based on $H_{0} = 75 \ km \ s^{-1} \ Mpc^{-1}$ and an
heliocentric radial velocity $V = 1100\ km \ s^{-1}$, which
implies that 1 arcsec = 0.07 kpc.

\begin{table}
\begin{minipage}[t]{\columnwidth}
\caption{General properties of NGC1947}             
\label{General}      
\centering
\renewcommand{\footnoterule}{}                       
\begin{tabular}{l l l}        
\hline\hline                 
Parameter & Value & Ref. \\    
\hline                        
   Morphological type & S0 peculiar &  NED\footnote{NASA/IPAC Extragalactic Database}\\       
   R.A. (J2000) & 05h26m47.6s & NED \\
   Dec. (J2000) & -63d45m36s & NED \\
   Helio. Radial Velocity & 1100 km/s & NED  \\
   Redshift & 0.0037 & NED     \\
   Distance & 15 Mpc &     \\
   Axial ratio$_{K}$ & 0.98 &     \\
   $(J-K)_{T}$ & 0.931 & 2MASS\footnote{2 Micron All Sky Survey}\\
   Total m$_{J}$ (mag) & 8.445 $\pm$ 0.02 & 2MASS \\
   Total m$_{K}$ (mag) & 7.510 $\pm$ 0.03 & 2MASS \\
   $M(HI)(M_{\bigodot})$ & $3.4 \times\ 10^{8}$ & \cite{Oos02}\\
   $M_{dust} (M_{\bigodot})$ & $15.3 \times\ 10^{5}$ & \cite{Oos02}\\
   $L_{K} (L_{\bigodot})$ & $2.65 \times\ 10^{11}$ & \\
   $M/L_{K} (M_{\bigodot}/L_{\bigodot})$ & 0.2 & \\

\hline                                   
\end{tabular}
\end{minipage}
\end{table}

\section{Observations and data reduction}\label{obs}

{\it Near-Infrared data} - NGC1947 belongs to a selected sample of
peculiar galaxies observed (December 2002) in the Near-Infrared
(NIR) J and K bands with the SofI infrared camera at the ESO-NTT
telescope. The field of view was $4.92 \times 4.92\ arcmin^2$ with
a pixel scale of 0.292 arcsec/pixel. Images were acquired in the
offsetting mode: a cycle was defined by several images on the
target, interspersed with sky frames and with an integration time
of 60 seconds; each object frame was taken with a small offset from
the galaxy center and the sky frames were taken before and after
each galaxy frame. More cycles were obtained in the K band than in
the J band, in order to have a better estimate of the background
level.  A total exposure time of 720 sec was obtained on the
target in the J band and of 2760 sec in the K band. The average
seeing during the observing time is about $FWHM \simeq 1.1$
arcsec.

\begin{figure*}
\centering
\includegraphics[width=15cm]{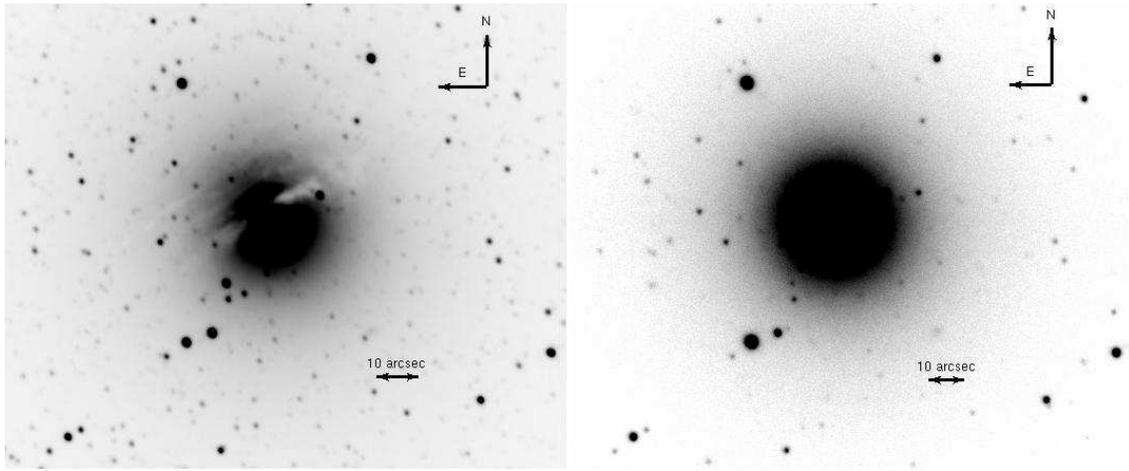}
\caption{Calibrated K (right panel) and V (left panel) images of
NGC1947} \label{N1947}
\end{figure*}

The data reduction was carried out using the CCDRED package in the
IRAF\footnote{IRAF is distributed by the National Optical
Astronomy Observatories, which is operated by the Associated
Universities for Research in Astronomy, Inc. under cooperative
agreement with the National Science Foundation.} ({\it Image
Reduction and Analysis Facility}) environment.  The main strategy
adopted for each data-set included dark subtraction\footnote{Bias
frame is included in the Dark frame.}, flatfielding correction,
sky subtraction and rejection of bad pixels.  Finally, all frames
were registered and co-added to form the final science frames.
Several standard stars, from \citet{Per98}, observed at the
beginning, middle and end of each observing night, were used to
transform instrumental magnitudes into the standard J and K band
systems.\\ The calibrated K band image of NGC1947 is shown in
Fig.~\ref{N1947} (right panel).

{\it Optical data} - CCD images of NGC1947 in the Johnson V and R
bands were extracted from the European Southern Observatory (ESO)
Public Archive. They had been obtained with the red arm of EMMI at
ESO-NTT telescope and the detector used for the observations
consisted of two MIT/LL CCDs, with a scale of 0.332 arcsec/pixel
(with a $2 \times 2$ binning) and a field of view of $22.7 \times
22.7$ $arcmin^2$.  The average seeing was $FWHM \sim 0.99 \pm
0.03$ arcsec for the V band, and $FWHM \sim 0.94 \pm 0.03$ for the
R band.  There were seven exposures available in the V band with
an integration time of $100 s$, and five exposures in the R band
with an integration time of $60 s$. The raw frames were
pre-processed using the standard IRAF tasks for de-biasing and
flat-fielding.  The flat-field was performed using normalized sky
flat.  Each CCD was reduced independently: the images were
dithered and shifted in order to obtain a complete image of both
the galaxy and its surroundings.  The frames of each CCDs were
combined, after de-biasing and flat-fielding, to filter out cosmic
ray hits.  Finally, we used the IRAF task IRMOSAIC to obtain a
mosaic of the images. Multiple exposures of identical fields were
aligned using several stars present in the images and then
co-added to produce a final median averaged image. An average
value of the background emission was estimated in several regions
of the sky far from the galaxy light.\\ In order to perform the
photometric calibration, several Landolt standard stars were
observed in the same bands.\\ The final, calibrated V band image
of NGC1947 is shown in Fig.~\ref{N1947} (left panel).\\ {\it
Spectroscopic data} - The spectroscopic data were extracted from
the European Southern Observatory (ESO) public
archive\footnote{http://archive.eso.org/}. They have been obtained
with the EMMI spectrograph at ESO-NTT in the RILD mode. The data,
consisting of four set of spectra, were acquired with a slit
1.0$''$ wide, using the GR$\sharp5$ grism with a dispersion of 5.5
nm mm$^{-1}$, corresponding to 0.83 \AA\ pxl$^{-1}$, in the
$380-700$ nm wavelength range. The spectra were acquired along
both the photometric major ($P.A.= 40^{\circ}$) and minor
($P.A.=130^{\circ}$) axis of NGC1947. The total integration time
of both spectra is 600 s; the average seeing turned out to be
$FWHM =1$ arcsec. A set of spectra of standard template $K0 III$
stars were also acquired with the same configuration.\\ Individual
frames were pre-reduced using the standard IRAF image processing
package; the wavelength calibration was made using the IRAF
TWODSPEC.LONGSLIT package and a set of He-Ar-Ne lamp spectra taken
for each observing night. Sky subtraction was performed with the
IRAF BACKGROUND package, by using a narrow region at both edges of
the slit where the galaxy contamination is minimum.  Finally, all
the exposures for both axis were co-added in the final median
averaged 2D spectra, which were used to derive the central
velocity dispersion and the rotation curve along both axes. The
velocity dispersion and the radial velocity were derived with the
Fourier Quotient Technique (\citealt{Sarg77}; \citealt{Bert84}),
using the standard stars as templates. One of the main source of
statistical and systematics errors is the template mismatching; so
we performed tests to verify if the standard star we used to
derive the kinematics of the galaxy was the best one. These tests
showed no template mismatching. In both spectra some extended
emission lines are well visible, in particular [OIII]
$\lambda5007$\AA, $H_{\alpha}$, [NII] $\lambda\lambda$6548,6583
and [SII] $\lambda\lambda$6717,6731. These lines allowed us to
measure the gas rotation curves also. To measure the rotation
curves, every record of each galaxy spectrum was fitted with a
Gaussian function on the emission line using the task SPLOT of
IRAF package.  The $H_{\alpha}$ line was not measured due to the
presence of a strong absorption component, for this reason only
the more extended [NII] $\lambda\ 6583$ line with a signal to
noise ratio $\geq 2$ was fitted. Taking into account the average
seeing and the EMMI pixel scale, the final stellar and gaseous
rotation curves were derived by adopting a binning factor of 3
along the spatial direction to obtain independent data points.
They will be presented in Fig.\ref{kin} and described in the
sec.\ref{kinem}. Due to the low signal-to-noise ratio (which is
about 20 in the center and decreases to about 10 towards the outer
regions), the kinematic profiles are about 4 times less extended
than the light profiles (Fig.\ref{ell_JK}).


\section{Photometry: morphology, light and color distribution}\label{prof}

The most prominent feature of NGC1947 is the extended dust-lane, which
crosses the galaxy center along the photometric minor axis.  Since the
dust optical depth decreases toward longer wavelengths, NIR photometry
is necessary to reduce as much as possible the dust absorption, that
strongly affects the starlight distribution in the galaxy. Thus, we
used near-infrared images (J and K bands) to easily identify the inner
structures of NGC1947 and accurate analyse the stellar light
distribution. On the other hand, optical images (in the V and R bands)
are used to derive the optical versus NIR color profiles and color
maps to study the structure of the dust-lane.

\begin{figure*}
   \centering
\includegraphics[width=15cm]{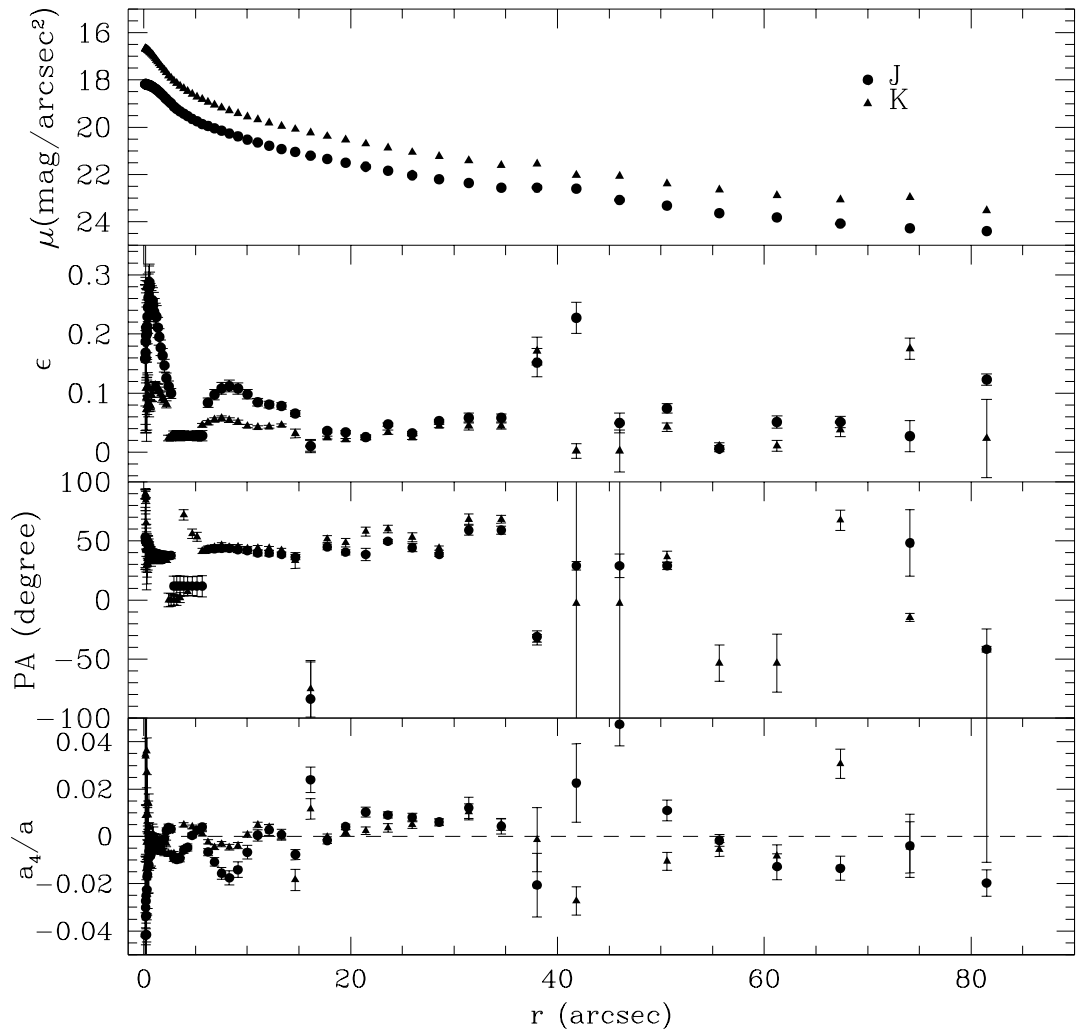}
      \caption{Ellipticity ($\epsilon$), Position Angle (P.A), mean surface brightness profile and shape parameter
      $a_{4}/a$ in the J and K bands.}
         \label{ell_JK}
   \end{figure*}

{\it Isophote analysis} - We used the IRAF-ELLIPSE task on the NIR
images to perform the isophotal analysis for NGC1947 and the results
are shown in Fig.\ref{ell_JK}. The average surface brightness extends
up to about 80 arcsec from the galaxy center (see Fig.\ref{ell_JK}); in
the K band, the half-light radius is $R_e = 24.4 \pm 2$ arcsec $\sim
1.86 \pm 0.15$ kpc. For a semi-major axis {\it r}, in the range $10
\le r \le 40$ arcsec, the ellipticity and P.A. are almost constant and
equal to 0.05 and $40^{\circ}$, respectively, that indicates that in
this regions the isophotes are almost round and co-axial. For $r \le
20$ arcsec the dust lane perturbs the ellipticity and P.A. profiles
and also the shape parameters $a_4/a$, which are shown in
Fig.\ref{ell_JK}.  This effect is even more evident in the J than in
the K band.  For $10 \le r \le 25$ arcsec, the shape parameters are
all consistent with zero, thus the isophotes do not significantly
deviate from purely elliptical shape, while at larger distances (out
to $\sim 40$ arcsec) they tend to be more disky (see
Fig.\ref{ell_JK}). In Tab.\ref{tab_mag} we give the integrated
magnitudes and average colors within two circular apertures, derived
for the NIR J and K bands, and optical V and R bands. The apertures
were chosen in order to make the comparison easier with the magnitudes
of NED.

\begin{figure*}
   \centering
   \includegraphics[width=6.6cm]{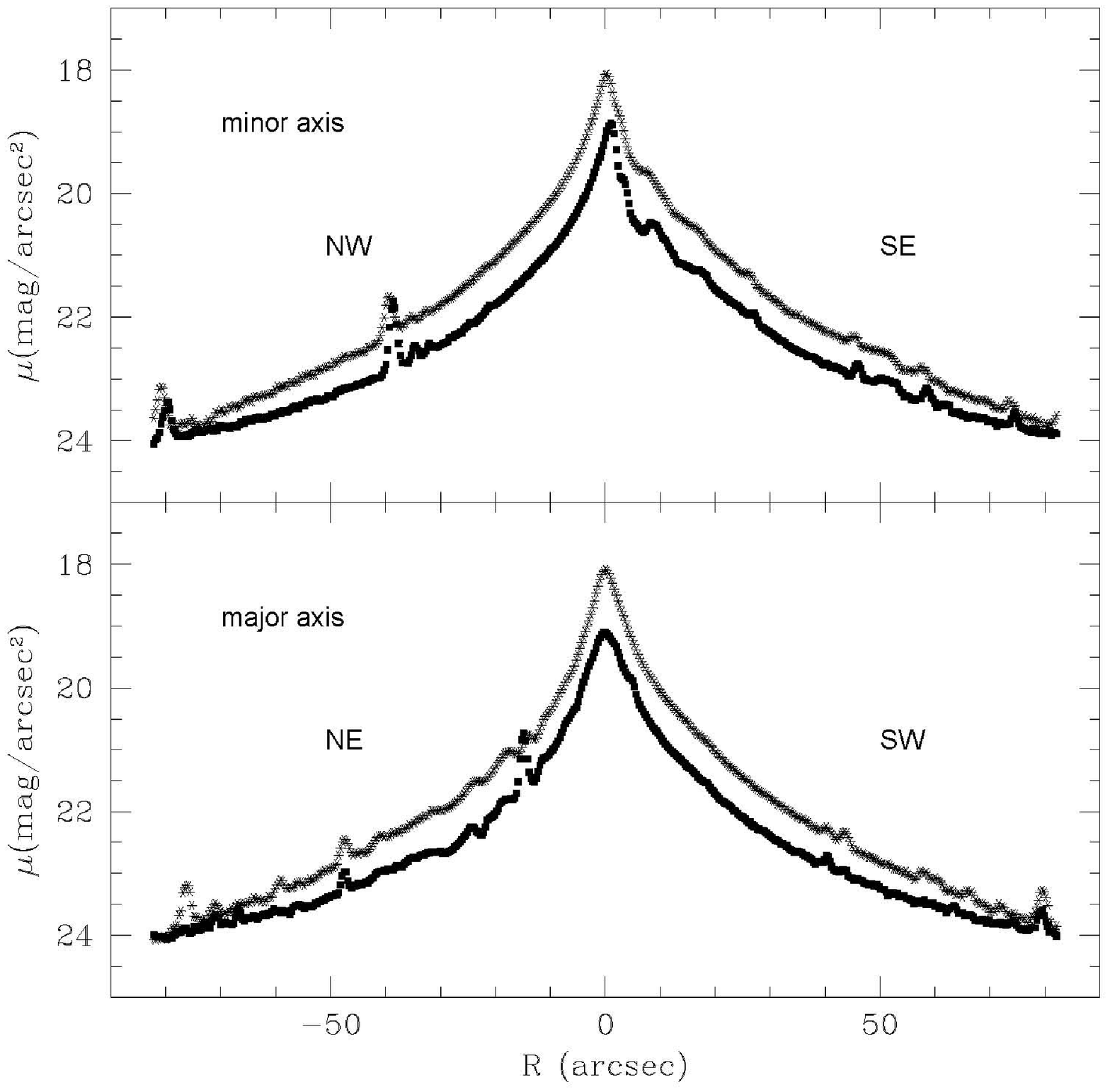} \hspace{1em}
   \includegraphics[width=6.6cm]{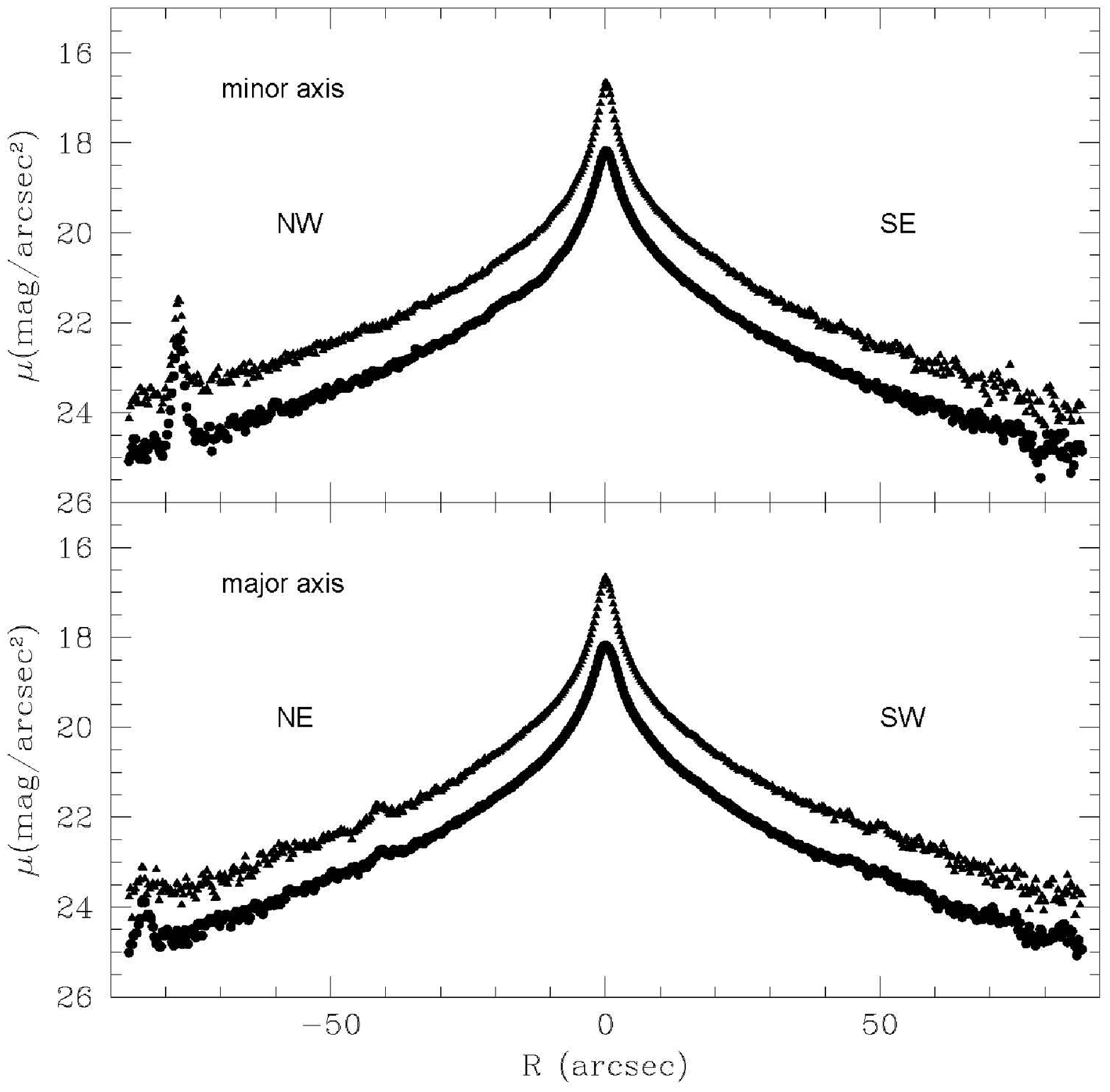} \caption{Light profiles
   along the minor (top-left panel) and major (bottom-left panel) axis
   of NGC1947 in the V (stars) and R band (triangles).  Light profiles
   along the minor (top-right panel) and major (bottom-right panel)
   axis of NGC1947 in the J (stars) and K band (triangles)}
\label{profili}
\end{figure*}

\begin{figure*}
  \centering
 \includegraphics[width=10cm]{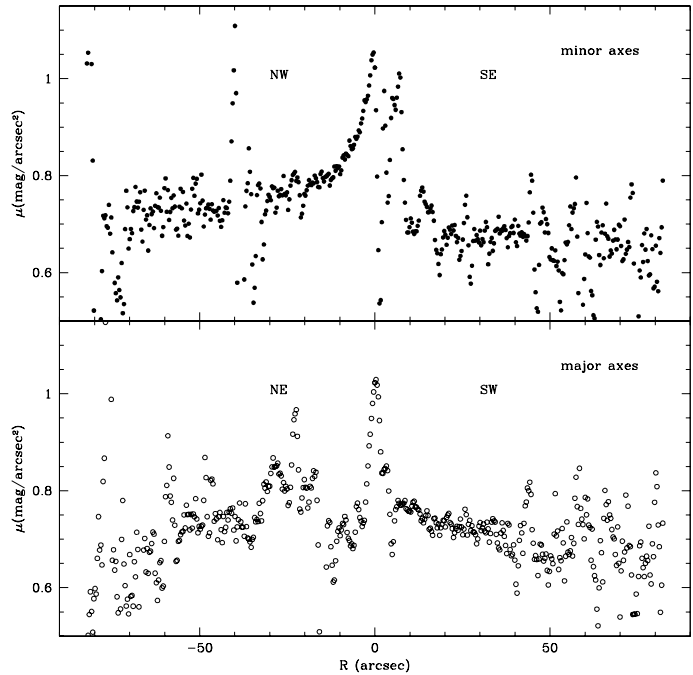}
      \caption{V-R color profiles along the minor (top panel) and major (bottom panel) axis of NGC1947.}
       \label{colorVR}
   \end{figure*}

\begin{figure*}
  \centering
    \begin{minipage}[c]{0.5\linewidth}
      \includegraphics[width=6.6cm]{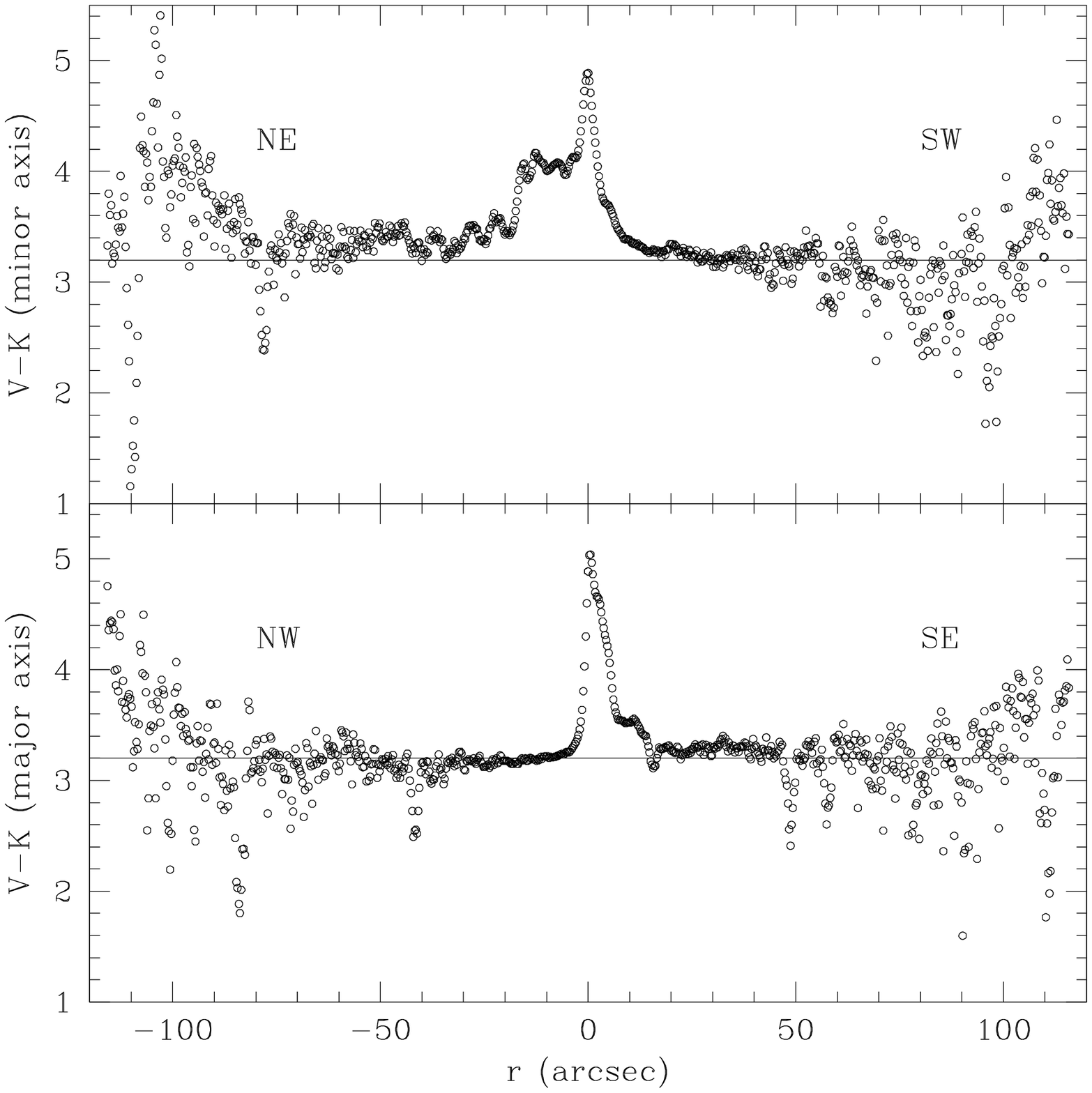}
    \end{minipage}\hfill
    \begin{minipage}[c]{0.5\linewidth}
     \includegraphics[width=7.5cm]{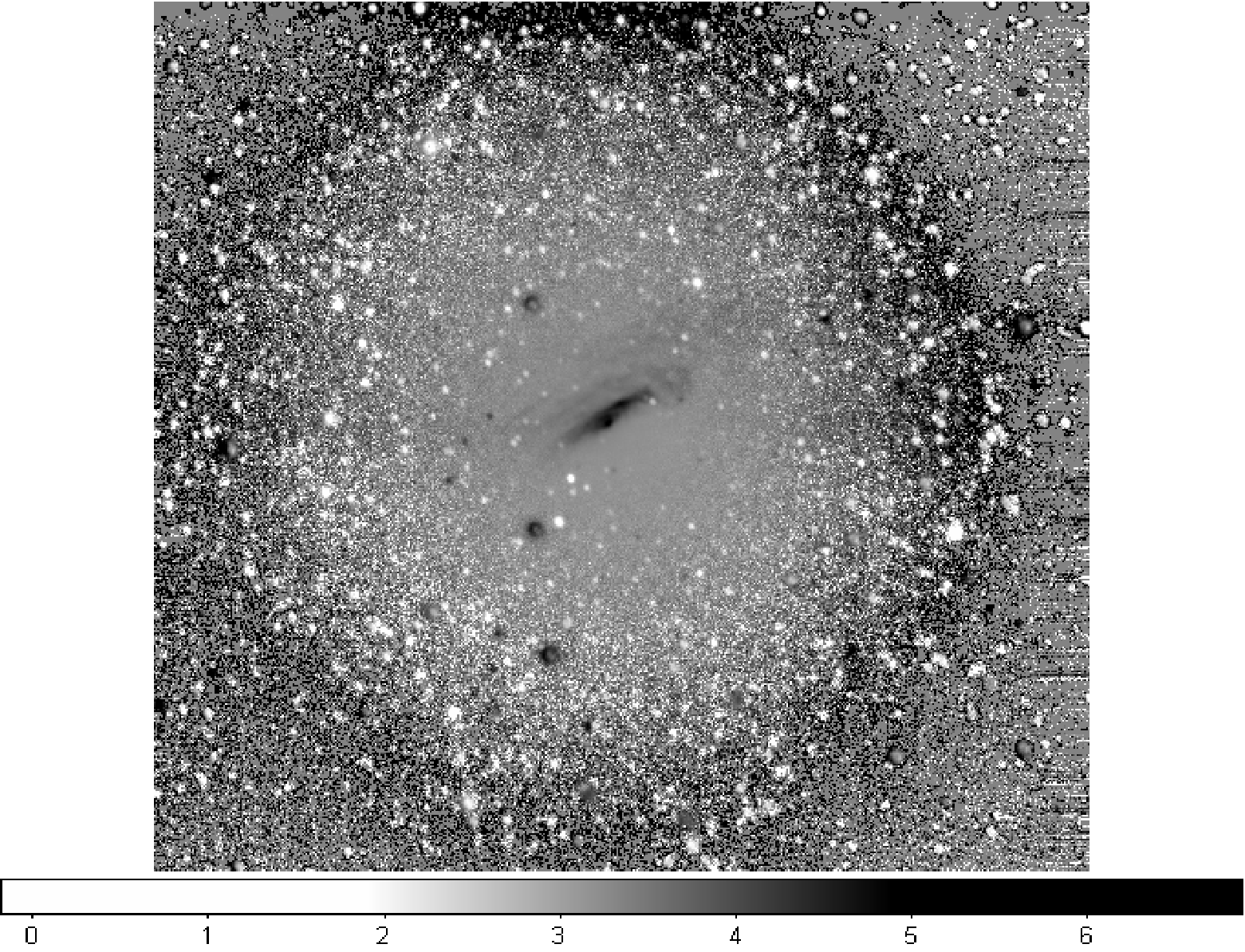}
    \end{minipage}
\caption{\emph{Left panel} - V-K color profiles along the major
(top panel) and minor (bottom panel) axis. The straight line
correspond to the average V-K color for $r \ge 10$ arcsec.
      \emph{Right panel} - V-K color map. The North is up, while the east is on the left of the image. Darker colors correspond to redder galaxy regions.}
       \label{VK}
\end{figure*}

\begin{figure*}
  \centering
    \begin{minipage}[c]{0.5\linewidth}
      \includegraphics[width=6.6cm]{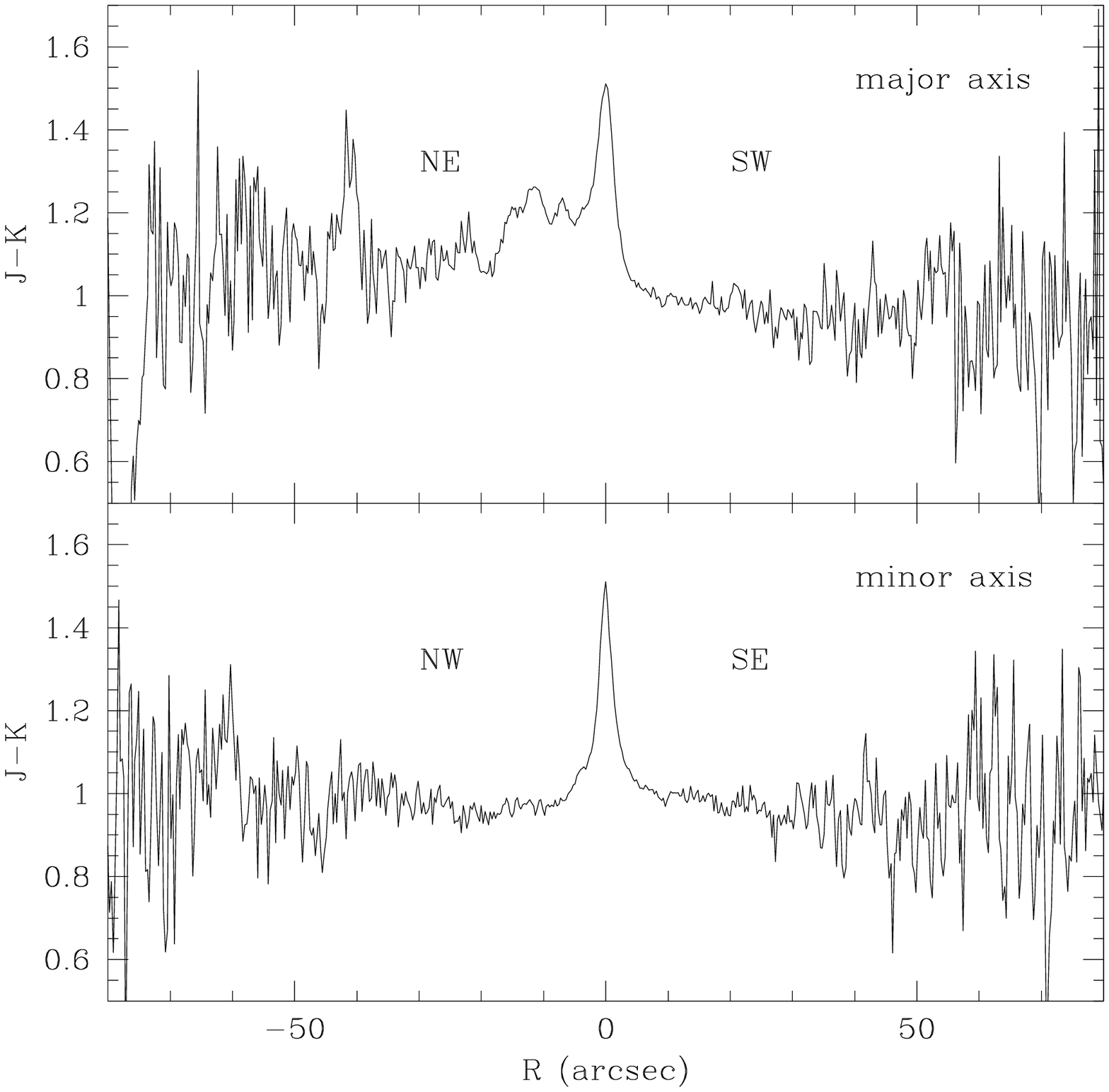}
    \end{minipage}\hfill
    \begin{minipage}[c]{0.5\linewidth}
     \includegraphics[width=7.5cm]{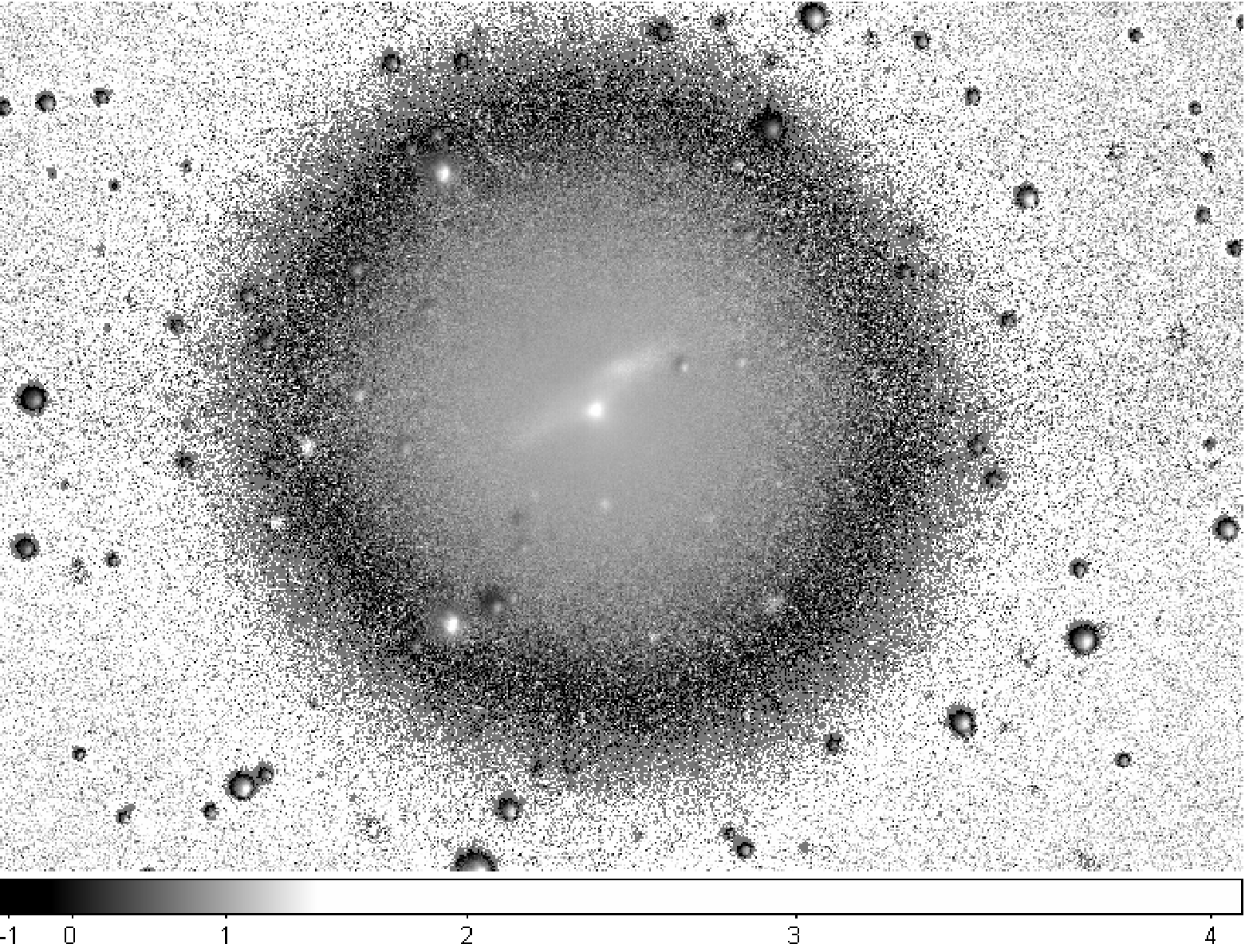}
    \end{minipage}
\caption{\emph{Left panel} - J-K color profiles along the major
(top panel) and minor (bottom panel) axis.
      \emph{Right panel} - J-K color map. The North is up, while the east is on the left of the image. Lighter colors correspond to redder galaxy regions.}
       \label{JK}
\end{figure*}

{\it Light and color distribution} - In Fig.\ref{profili} we show
the light profiles along the major ($P.A.=40^{\circ}$) and minor
($P.A.=130^{\circ}$) axis in the NIR and in the optical bands
respectively. The light distributions in the optical V and R bands
are very much perturbed by the presence of the dust-lane, in the
regions between 10 and 30 arcsec on the NE side of the major axis,
and for $r < 20$ arcsec on the SE side of the minor axis. Except
for this, neither NIR nor optical profiles show any significant
feature.\\ We have derived the $V-R$ (Fig.\ref{colorVR}), $J-K$
(Fig.\ref{JK}, left panels) and $V-K$ (Fig.\ref{VK}, left panels)
color profiles along both photometric axes of NGC1947, and the
2-dimensional $V-K$ and $J-K$ color maps (Fig.\ref{VK} and
Fig.\ref{JK} right panels, respectively). On average, the central
regions of the galaxy have redder colors, with a maximum value of
$V-R=1.05$, $J-K=1.5$ and $V-K=5$. At larger radii the colors are
bluer. In all color profiles, on the NE side of the major axis,
between 10 to 30 arcsec, double red peaks are observed, at r=18
arcsec and r=30 arcsec where V-R=0.85 and V-K=4: such features are
due to the ring-like structure of the dust-lane, already suggested
by \citet{Bert92} and described in Sec.\ref{intro}, which stands
out very clearly in the 2-dimensional V-K color map
(Fig.\ref{VK}). The dust-lane perturbs only the SE side of the
minor axis color profile, within 10 arcsec from the center: this
suggests that the dust-lane is slightly misaligned with respect to
the photometric minor axis. This is also confirmed by the
2-dimensional J-K color map (see Fig.\ref{JK}, right panel): with
respect to the nucleus, which is very red, there are two ``thin''
and redder structures which do not lie on the same direction. In
particular, the one on the NW side is on a parallel direction
towards North. The full extension of the inner dust-lane is about
27 arcsec.

{\it Stellar population ages} - We derived the integrated V-K and
J-K colors in two circular apertures: one including the nuclear
regions of the galaxy inside a radius of 1 kpc $\sim 0.54 R_e$,
where $V-K = 3.2$ and $J-K=1.02$, and one out to $1.5 R_e \sim
2.85$ kpc, where $V-K = 2.84$ and $J-K=0.99$.  The stellar
population synthesis model by \citet{Bruz03} were used to
reproduce the integrated colors in these regions, in order to
derive an estimate of the stellar population ages. We adopted a
star formation history with an exponentially decreasing rate, that
produces a reasonable fit of the photometric properties of
early-type galaxies in the local Universe. It has the following
analytical expression: $SFR(t)=\frac{1}{\tau} \exp{(- t/ \tau)}$,
where the $\tau$ parameter quantifies the ``time scale'' when the
star formation was most efficient. Adopting $\tau=1$ Gyr, the
correspondent evolutionary tracks were derived for both solar
metallicity, $Z=0.02$, and for a higher value, $Z=0.05$; both
values were assumed constant with age. In every model it has been
assumed that stars form according to the \citet{Sal55} IMF, in the
range from $0.1$ to $125 M_\odot$.  To account for the V-K and J-K
colors the best model is that obtained for $Z=0.05$, from which we
derived an age of about 3.9 Gyrs for the inner region and of about
2.3 Gyrs for the outer one. These estimates have to be considered
as an upper limit to the last burst of star formation. These
constraints will turn to be fundamental in the discussion on the
formation scenarios (Sec.\ref{disc}).

\subsection{2-Dimensional model of the light distribution}\label{2D}

\begin{figure*}
  \centering
 \includegraphics[width=10cm]{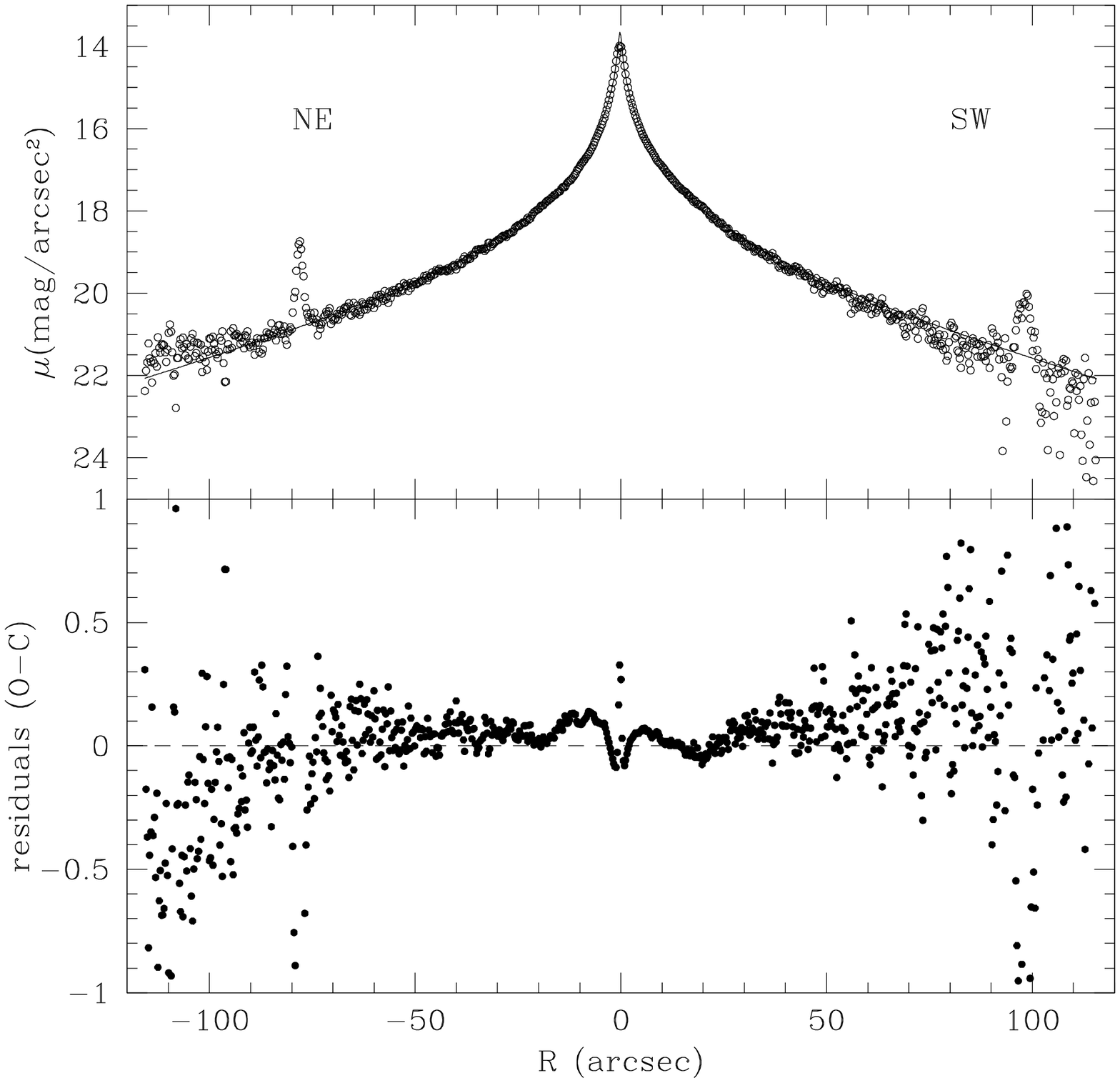}
      \caption{Top panel - 2-D fit of NGC1947 light distribution. The observed light profile along the major axis, is compared
      with those derived by the fit (continuous line).
   Bottom panel - Residuals between the observed and the fitted light profiles.}
       \label{prof_mj}
   \end{figure*}

\begin{figure*}
  \centering
 \includegraphics[width=10cm]{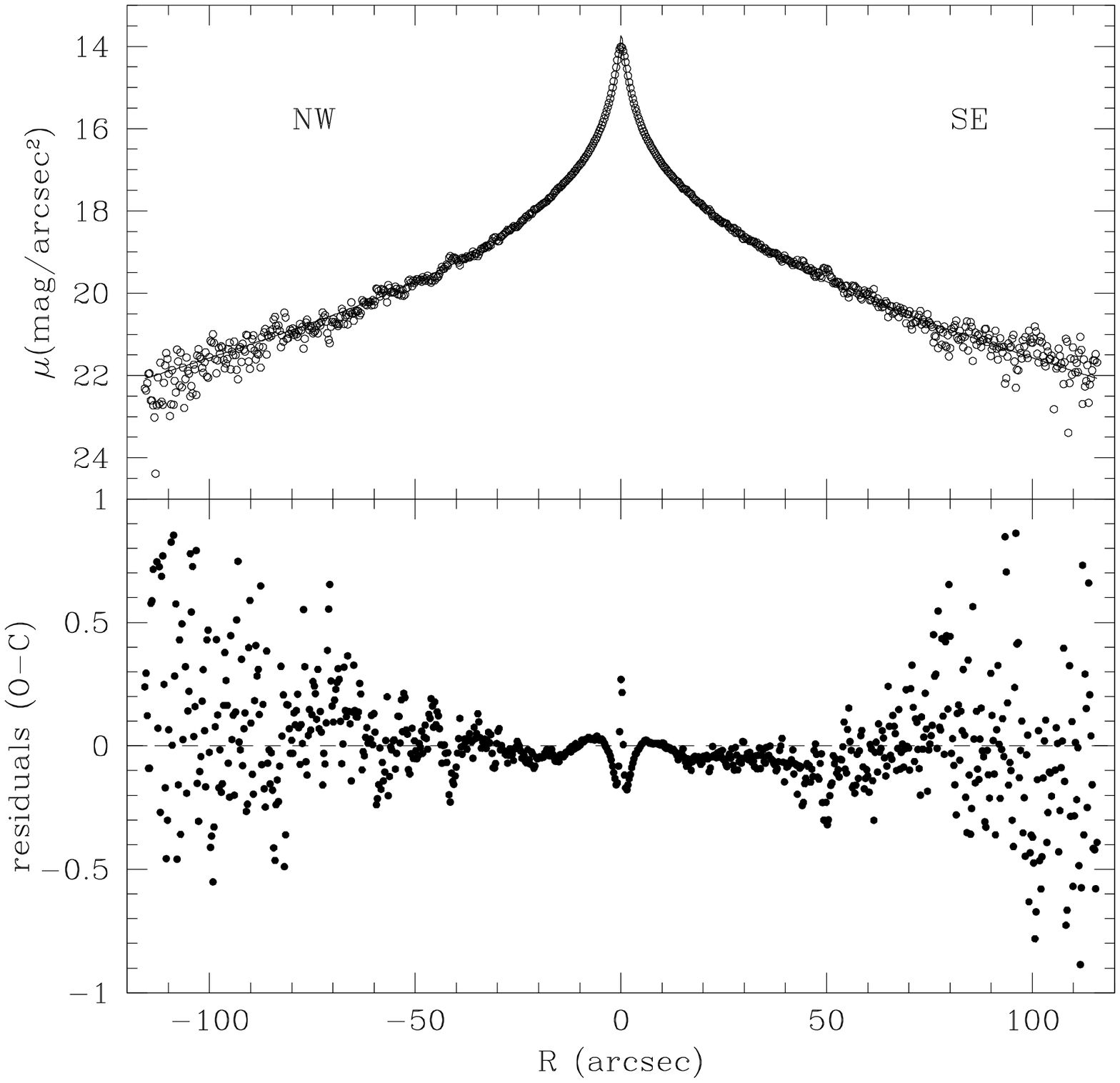}
      \caption{Top panel - 2-D fit of NGC1947 light distribution. The observed light profile along the minor axis, is compared
      with those derived by the fit (continuous line).
   Bottom panel - Residuals between the observed and the fitted light profiles.}
       \label{prof_mn}
   \end{figure*}

We performed a 2-dimensional model of the light distribution in
the K band, where the effect of dust absorption is weaker, and
stars are accurately masked. To this aim, we used the GALFIT task
\citep{Peng02}. As a first attempt and according to the
morphological classification given for NGC1947 in the previous
work (see Tab.\ref{General}) as an Elliptical, the galaxy light is
modelled through a single Sersic law \citep{Ser68}: the fit is
very poor, suggesting the presence of an additional component in
the outer regions. Therefore, we adopted a double-component model
as the superposition of a Sersic law \citep{Ser68} plus an
exponential profile \citep{Free70}. The fit improves considerably;
the fitted light profiles for the major and minor axis of NGC1947
are shown in Fig.\ref{prof_mj} and Fig.\ref{prof_mn}. The
structural parameters derived from the fit are: the bulge total
magnitude $m_{b}=17.7\pm 0.01\ mag$, the bulge scale length
$r_{e}=74.0\pm 0.7$ arcsec, the disk total magnitude
$m_{d}=17.0\pm 0.02\ mag$, the disk scale length $r_{h}=121.1\pm
0.3$ arcsec and the Sersic exponent $n=2.83\pm 0.01$, leading to a
{\bf bulge total luminosity-to-disk total luminosity ratio},
$L_{B}/L_{D}\simeq 0.43$.

\begin{figure*}
  \centering
 \includegraphics[width=13cm]{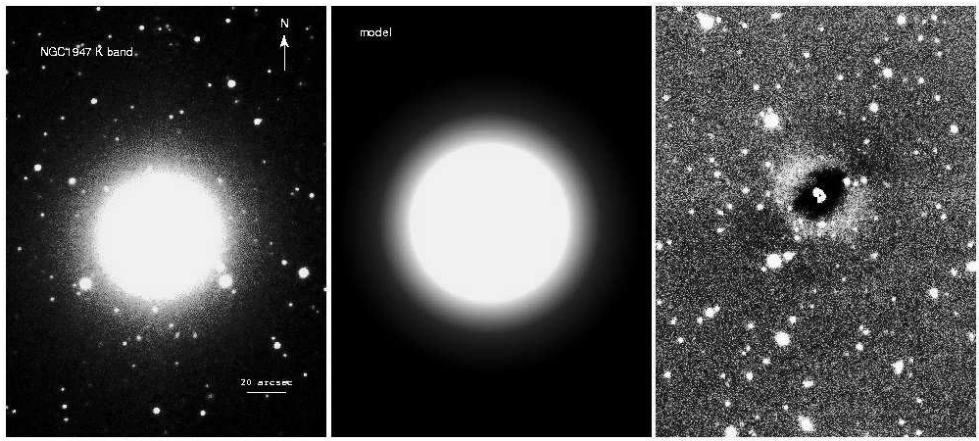}
      \caption{2D fit of NGC1947. Left panel - K band image of NGC1947. Middle panel - Model of the galaxy. Right panel - Residual image.}
       \label{2Dmod}
   \end{figure*}
\begin{figure*}
  \centering
 \includegraphics[width=10cm]{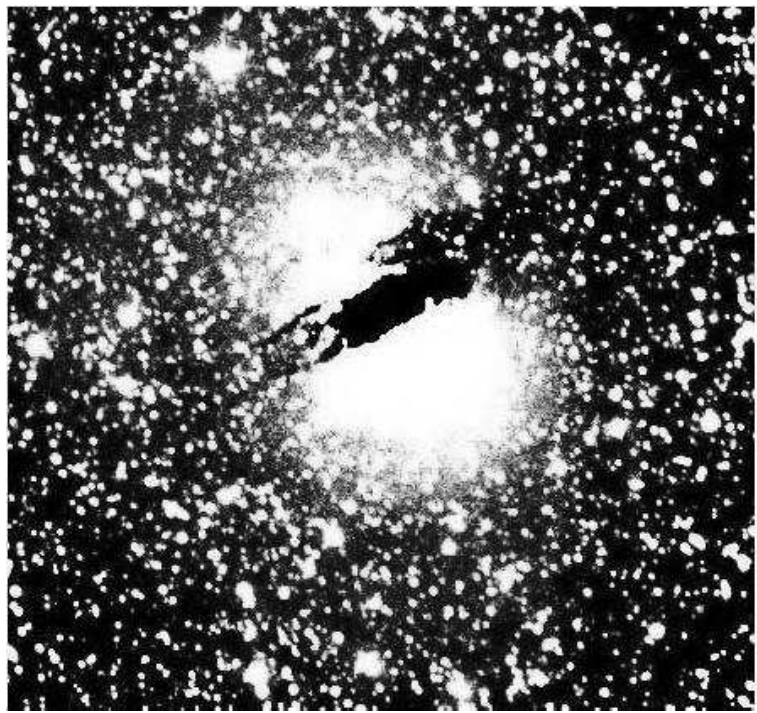}
      \caption{Residuals of the subtraction of the K band model to the V band image.}
       \label{res_v}
   \end{figure*}

Along both axes, inside a 30 arcsec radius from the galaxy center, the
residuals reflect the presence of the dust-lane, which is more
pronounced along the NE region of the major axis, where the galaxy
appears to be less luminous than the model.  Such an effect is much more
clear in the residual image (see Fig.\ref{2Dmod}) obtained by
subtracting the 2D model to the galaxy. In order to analyze the
structure of the dust-lane in NGC1947, we used the 2D model obtained
in the K band, which is less perturbed by dust absorption, to match
the light distribution in the V band. To this aim, the V band image of
NGC1947 was scaled to the K band, by accounting for the average V-K
color (see Sec.\ref{prof}). In the residuals (Fig.\ref{res_v}) two
main features stand out very clearly: \emph{i}) the concentric
ring-like structure of the dust-lane, superposed to \emph{ii}) a
bright region (within 20 arcsec from the center) which is more
luminous on the SW side with respect to the NE side.  The existence of
such luminous "counterpart" is reliable due to the fit interpolation,
which is an average of the whole galaxy light (see Fig.\ref{prof_mj}
and Fig.\ref{prof_mn}): this is also confirmed by the absence in the
V-K color profile (see Fig.\ref{VK}) of any bluer component on the SW
region, which should have appeared in the V-band residual image
respect to the K band one. By this analysis, we estimate that the
dust-lane extends by 49.6 arcsec from SE to NW, with a thickness of
about 23 arcsec.

\begin{table*}
\caption{Magnitudes and Colors for NGC1947 in circular aperture.}
\label{tab_mag}      
\centering                     
\begin{minipage}{140mm}
\begin{tabular}{lcccccc}        
\hline\hline                 
aperture radius&$m_J$& $m_K$ & $m_V$ & $m_R$ & J-K & V-R\\
~ (arcsec) &$\pm$ 0.05 &$\pm$ 0.04 & $\pm$ 0.01 & $\pm$ 0.01 & $\pm$ 0.09& \\
\hline
15.7 & 10.44 & 9.35& 13.20 & 11.81 & 1.09 &1.39 \\
81.2 & 8.91& 7.90& 11.26 & 9.99 & 1.01 &1.27\\
\hline
\end{tabular}
\end{minipage}
\end{table*}


\section{Stellar and gaseous kinematics along the principal photometric axes
of NGC~1947}\label{kinem}

\begin{figure*}
   \centering
\includegraphics[width=10cm]{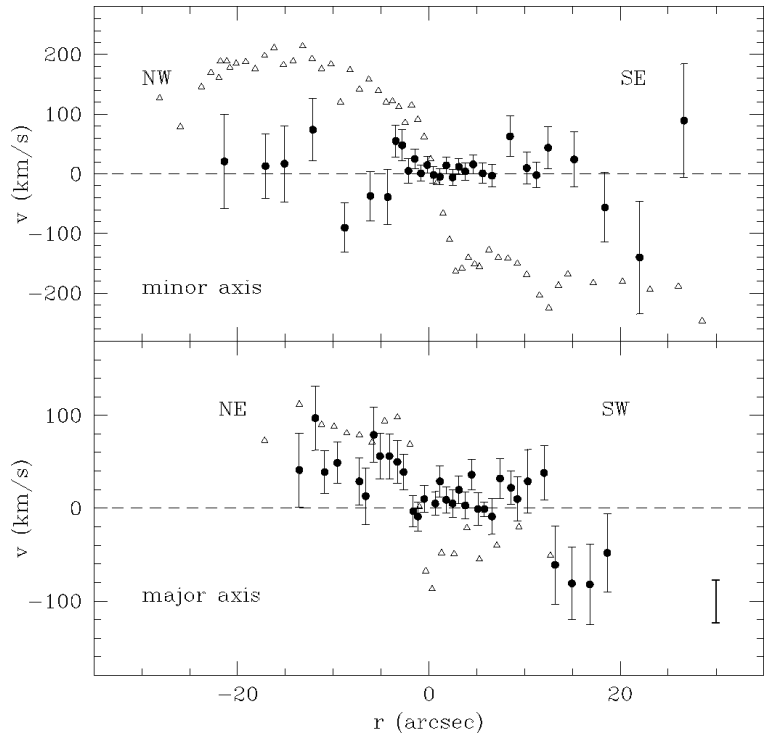}
      \caption{Stellar (filled points) and gaseous (triangles)
      rotation curves along the major (bottom panel) and minor axis
      (top panel) of NGC1947. The dashed line is the velocity of the
      galaxy center. In the bottom-right corner is plotted the error bar that
      represents the mean error on the velocity of the gas.}  \label{kin}
\end{figure*}

{\it Major axis} - The stellar kinematics is more extended on the SW
side with respect to the NE and it is measured out to $20'' \sim 1.4$
kpc (Fig.~\ref{kin}, bottom panel), which corresponds to a surface
brightness of $\sim 21.7$ mag arcsec$^{-2}$ in the V band.  The
rotation curve is not symmetric with respect to the galaxy center: in
the range $2''< r <14''$ NE, rotation increases, up to a maximum value
between 55 and 80 km/s, to be compared with the corresponding distance
range on the SW side, where almost no rotation is measured. The
kinematics along NE side, for $r > 6''$, is strongly perturbed by the
dust-lane, as suggested by the V-K color profiles (Fig.\ref{VK}).\\
The velocity dispersion remains almost constant within the
uncertainties at an average value of $\sim 133 \pm 31$ km/s, out to 10
arcsec from the center. Both the rotation velocity profile and the
average velocity dispersion turns out to be consistent with those
published by
\citet{Bert92}.\\ The gas component too shows some rotation along
this direction, and the rotation curve has the same extension as the
stellar rotation curves (see Fig.~\ref{kin}): it shows a very steep
inner gradient, reaching a maximum velocity of $\sim 100$ km/s on
the NE side and $\sim 80$ km/s on SW side, between $2''$ and $4''$
from the center. Velocities remain almost on these values up to the
last measured point. These data (both for emission and absorption) are
consistent with, and extend those of \citet{Mol82}. However,
the measured rotation for the gas was not so evident in the data of
\citet{Mol82}.

{\it Minor axis} - The stellar kinematics is measured out to the same
distance from the galaxy center, i.e. $20''$ (Fig.~\ref{kin}, top
panel). On average, stars do not rotate and the velocity dispersion
has the mean weighted value of $\sim 146 \pm 37$ km/s within $5''$
from the center.\\ The gas rotation curve is more extended (up to
about $30''$) and better resolved with respect to the stellar rotation
curves. Along the major axis, the gas shows a very steep inner gradient,
reaching a maximum velocity of $\sim 200$ km/s between $10''$ and
$20''$ from the center. The full extension of the gas rotation curve
is of about 60 arcsec, which is about 2 times wider than the
dust-lane (see Sec. \ref{prof}).\\
The gas velocity curve presented in this work, even though more
extended on the SE side, is consistent with the one published by
\citet{Mol82}.\\ The observed kinematics led us to conclude that the
minor axis is the principal rotation plane of the gaseous disk.
Moreover, for this component some rotation is observed even in the
direction of the major axis: this implies that the disk is warping
towards the inner regions \citep{Arna93} and suggests that this
component has not reached a stable configuration yet. The average
velocity dispersion turns out to be consistent also with values
derived by \citet{Car93}, according to the kinematics derived along
two intermediate directions.

\begin{figure*}
   \centering
\includegraphics[width=10cm]{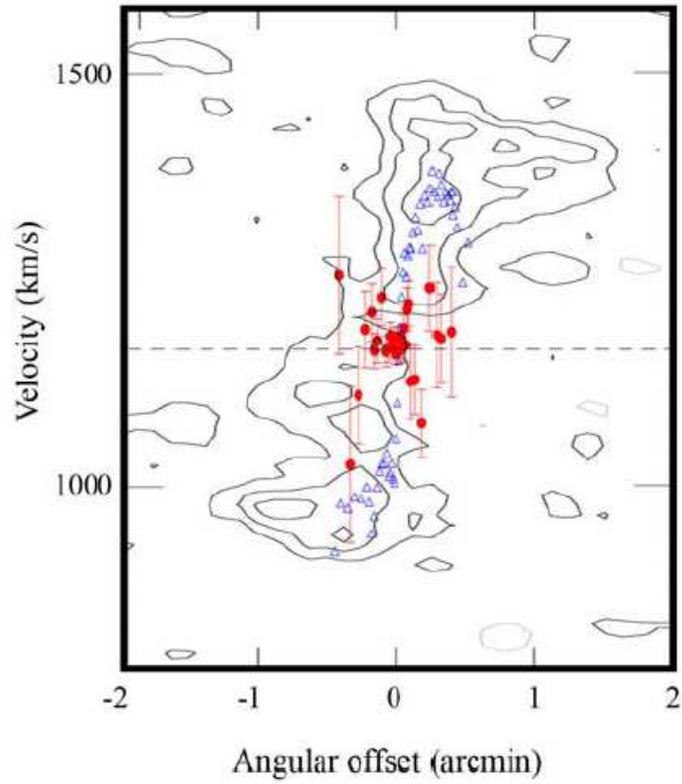}
      \caption{Stellar (filled points) and gaseous (triangles)
      rotation curves along the minor axis of NGC1947 superposed on
      the position-velocity plot of the HI along the $P.A.=127\circ$
      presented by \citet{Oos02}.}  \label{HI}
\end{figure*}

\begin{figure*}
  \centering
 \includegraphics[width=10cm]{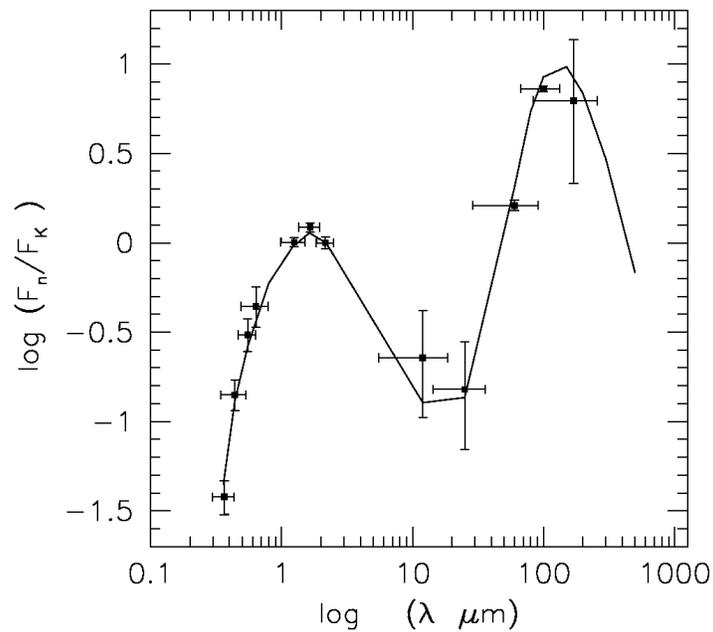}
      \caption{Continuous line shows the prediction of our model (see
      text); filled squares are  V, J, and K integrated data given in sec. \ref{prof}, the other data
      are from NED, in particular IRAS data at 12, 25, 60 100
      $\mu$m and ISOPHOT data at 170 $\mu$m of \citet{Sti04}.}
       \label{sed}
   \end{figure*}

In Fig. \ref{HI} we show the stellar and gaseous rotation curves along
the minor axis of NGC1947 superposed on the position-velocity plot
along the $P.A.=127^{\circ}$ of the HI distribution presented by
\citet{Oos02}. The HI distribution is more extended than the stellar
rotation curve: the HI extends out to about 90 arcsec, while the
stellar rotation curve extends up to about 20 arcsec. The HI and gas in
the dust-lane rotates in the same sense and with the same amplitude
(see Fig.\ref{HI}): this confirms that the HI is in a warped rotating
disk associated with the dust-lane, as already suggested by
\citet{Oos02}.


\section{Discussion and conclusions}\label{disc}

In the present work we performed a detailed analysis of the
structure of the minor-axis dust-lane galaxy NGC1947, and in what
follows we will discuss the implications of our findings on the
formation and evolution history of this peculiar object. In
particular, we will discuss how the observed properties for
NGC1947 presented in this work, compare with the predictions from
different formation scenarios for such kind of systems. The main
observed properties for NGC1947 that the most likely formation
scenario has to account for are {\it i)} the high-inclined ring of
gas and dust which rotates in the perpendicular direction with
respect to the stars in the major axies of the S0-like host
galaxy, and {\it ii)} the photometry and kinematics observed for
the two decoupled components (stars and gas), in particular, the
warp observed for the gas towards the inner regions (sec.
\ref{kinem}), which suggests that the inner rings of gas has not
settled yet in the equilibrium plane; {\it iii)} the estimated
timescale of the formation mechanism has to be consistent with the
age estimates of the different stellar populations.\\ The presence
of two components with different angular momenta, gas and dust
along the minor axis and stars along the major axis, suggests that
NGC1947 could not be the result of a single protogalactic cloud
collapse and that it may have experienced an interaction event. In
this framework, the possible formation scenarios which may account
for a galaxy with high inclined ring/disk of gas, stars and dust
(discussed in the Sec.\ref{intro}) are {\it i)} the tidal
accretion of gas by outside, and {\it ii)} the unequal-mass
merging of two disk galaxies. In the following we will
discriminate between these two scenarios for NGC1947. Furthermore,
we will discuss how Smoothed Particles Hydrodynamic (SPH)
simulations of isolated collapsing triaxial systems, presented by
\citet{MC03}, may account for peculiar galaxies like NGC1947.

- \emph{Tidal accretion} -

\begin{figure*}
  \centering
 \includegraphics[width=10cm]{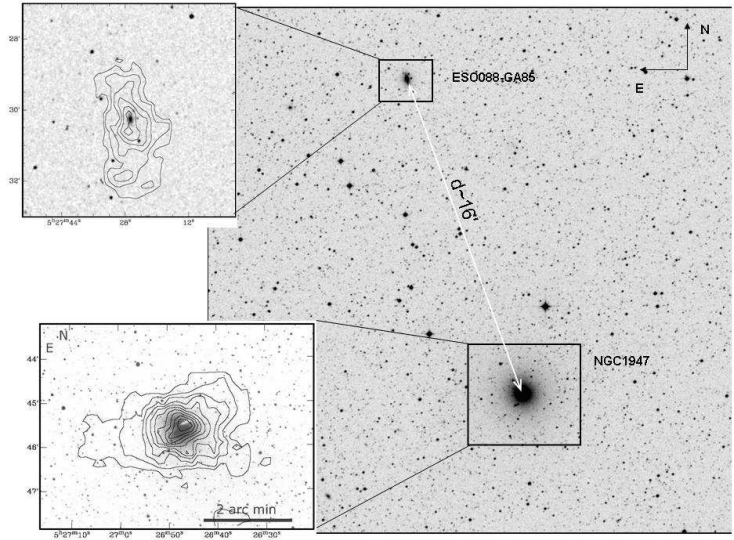}
      \caption{Image of a field of 25 arcmin around NGC1947. The two boxes on the left show the HI distribution
      for NGC1947 and ES0 085-GA088 published by \citet{Oos02}.}
       \label{campo}
   \end{figure*}

The features observed in NGC1947 lead to the possibility of
considering the galaxy as a tidal accretion remnant. The tidal
accretion scenario accounts for a large variety of polar
structures; many simulations show that the accretion produces
warped rings and transient features (see \citealt{Bour03}), with
processes similar to those leading to the formation of Polar Ring
Galaxies (PRGs). Differently from PRGs, in the polar ring of
NGC1947 there are not detected stars associated to the gas and
dust. Such Interstellar Medium (ISM) could be accreted by a
gas-rich donor galaxy during a close passage of a pre-existing S0
galaxy near this companion galaxy. In order to consider the
accretion hypothesis, we studied the field around NGC1947 (shown
in Fig.\ref{campo}) to see if there are any objects from which the
galaxy could have accreted material: inside a radius of about 5
times the diameter of the galaxy, which correspond to about 16
arcmin, (as suggested by \citealt{Broc97}) there is the early-type
spiral galaxy ESO~085-GA088, which is at the same redshift of
NGC1947, with a radial velocity of 1164 km/s. As discussed by
\citet{Oos02}, NGC1947 shows a warped HI disk at the same position
angle as the dust lane, while the HI emission of ESO 085-GA088
extends in a direction perpendicular to that of NGC1947: this is
the orbital configuration needed to form a polar ring (of gas)
through an accretion event \citep{Bour03}. These observed features
may suggest that the donor galaxy could be ESO~085-GA088: in fact,
taking into account the linear relative distance of about $16'
\sim 70$ kpc, and the relative velocities of the two systems, we
estimate the time at which the interaction happened, to be about 1
Gyr ago. Such value estimated in the case of a parabolic orbit has
to be considered a lower limit. In the case of a parabolic orbit a
longer time is expected, which should be consistent with the upper
limit to the last burst of star formation in NGC1947 (estimated in
Sec. \ref{prof}).

- \emph{Merging} -

Unequal-mass merging of two disk galaxies may build a spheroidal
galaxy with an high inclined ring/disk of gas, stars and dust. In
order to test such formation mechanism, we will compare the
observed properties with those expected from simulations performed
by \citet{Bour05}. To do so, we computed the following
quantities:
\begin{itemize}
\item the fitted light profile for this galaxy is consistent with
    a Sersic plus exponential luminosity profile, with bulge/disk=0.43
    and bulge/total=0.57 (see Sec.\ref{prof});

\item according to \citet{Bour05}, we evaluated the isophotal
shape parameters of the galaxy, in the range $[0.55R_{25th},
R_{25th}]$, which is the range between the bulge and the disk
optical radius. In this range the ellipticity is almost constant,
$\sim 0.11$, and the $a_{4}\simeq - 0.004$ suggests a bulge with a
moderate boxiness;

\item we have taken into account the global kinematical properties
of NGC1947 by deriving the $v/\sigma$ ratio along the major axis:
in the range $[0.55R_{25th},R_{25th}]$ $v/\sigma \sim 0.15$, while
the ratio $v_{max}/\sigma_{0}$ is about 0.62.
\end{itemize}

Such quantities have been compared with those predicted by
numerical simulations for galaxy mergers with various mass ratios
(see Tab.2 in \citealt{Bour05}): we found that they turn to be
consistent with those expected for a merging with initial mass
ratios 3:1 or 4:1 (\citealt{Bour05}). This is a mass ratio for
which \citealt{Bour05} found stable and long lived polar or
high-inclined rings of gas, returning to the galaxy along tidal
tails, and equatorial rings, that appears as "dust lanes" when
they are seen edge-on. Such unequal-mass merger has been proven to
be the most likely scenario for another peculiar galaxy: the
double-ring ESO~474-G26 \citep{Res05}.

\begin{figure*} \centering \includegraphics[width=10cm]{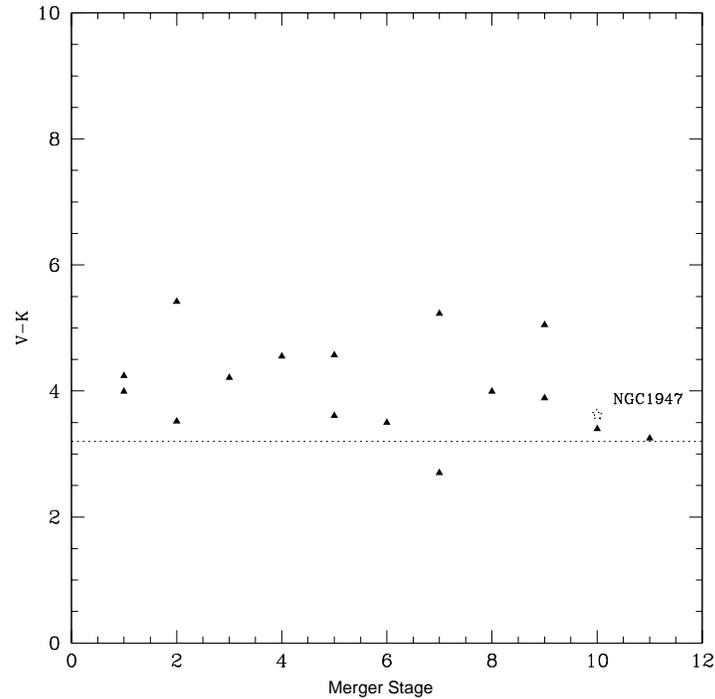}
 \caption{V-K color measured within 1kpc apertures as a function of
 the Toomre sequence merger stage.  The dotted horizontal line indicates the
 V-K color for a SSP $10^{10}yr$ population of solar metallicity,
 which provides a reasonable fit to the measured colors for early-type
 galaxies.}  \label{v_k}
\end{figure*}

\begin{figure*}
  \centering \includegraphics[width=10cm]{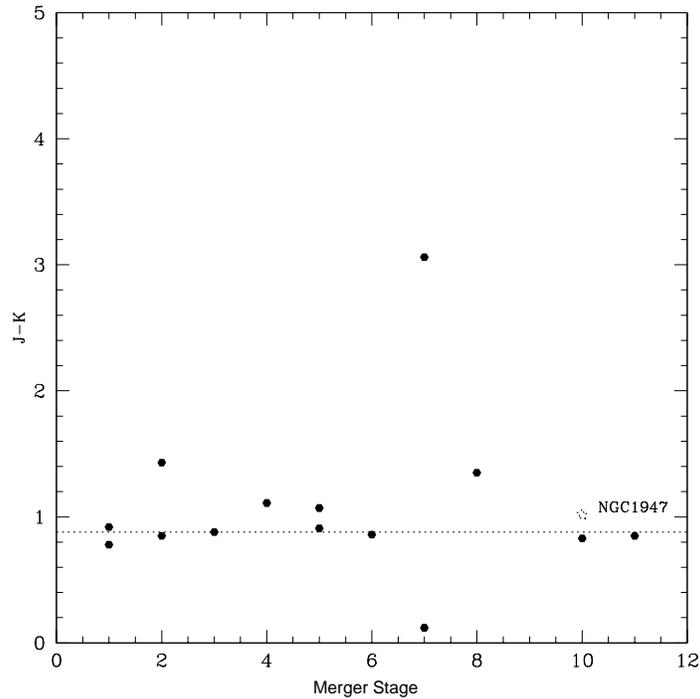} \caption{J-K
  color measured within 1kpc apertures as a function of the Toomre
  sequence merger stage.  The dotted horizontal line indicates the J-K color
  for a SSP $10^{10}yr$ population of solar metallicity.}  \label{j_k}
  \end{figure*}

In the hypothesis that NGC1947 may be the result of a merging
event, we compare the observed properties of NGC1947 with those of
the galaxies that have certainly experienced a merging, such as
the galaxies belonging to the Toomre Sequence (hereafter TS,
\citealt{Too77}).\\
The first step was to place NGC1947 on the TS, to determine at
which merger state we are looking at. To this aim, we studied the
position of NGC1947 on the plot of the total K band magnitude
versus the magnitude within a radius of 1 kpc for the galaxies
belonging to the TS, performed by \citet{Rossa07} (see Fig.25 in
that paper). Thus, for NGC1947 we estimated the total K band
magnitude and the V, J and K magnitudes within a radius of 1kpc,
which are respectively $M_{K} = -23.3$ mag, $m_{V} = 12.37$ mag,
$m_{J} = 9.79$ mag and $m_{K} = 8.74$ mag. The nucleus of NGC1947
lies in the region between the nuclei of the late-stage mergers
and the normal early-type galaxies. This leads to the reasonably
conclusion that NGC1947 could be considered a very late-stage
merger remnant. As a second step, we compared the V-K and J-K
nuclei colors obtained by \citet{Rossa07} for the galaxies
belonging to the TS with those derived for NGC1947 (see
Sec.\ref{prof}), by classifying it as late-stage merger
(Fig.\ref{v_k} and \ref{j_k}). The J-K colors for NGC1947 are
consistent with those of TS galaxies, while in the V-K color plot
NGC1947 turns to be bluer relative to the TS galaxies: since no
clumps of star forming galaxies are observed in NGC1947, this
difference could reasonably due to the presence of a larger amount
of dust in the TS galaxies.

- \emph{Formation of NGC1947 through SPH simulations} -

We derive a suitable match of the dynamical and photometric properties
of NGC1947 with a SPH simulation: \citet{MC03} implemented SPH
simulations with chemo-photometric evolutionary population synthesis
models providing the SED from ultraviolet to near-IR wavelengths. The
simulation includes self-gravity of gas, stars and dark matter (DM),
radiative cooling, hydrodynamical pressure, shock heating, artificial
viscosity, star formation and feedback from evolved stars and type II
SNe, as described in \citet{MC03}.  The starting point is a triaxial
collapsing system initially composed of 30000 particles, 15000 of gas
and 15000 of DM with a relative mass ratio of 0.01. The total mass, M,
of the collapsing system is 10$^{13}M\odot$. The system is built up as
described by \citet{MC03}, i.e., with a spin parameter, $\lambda$,
given by $|{\bf {J}}||E|^{0.5}/(GM^{0.5})$, where E is the total
energy, J the total angular momentum and G the gravitational constant,
equal to 0.06, and it is aligned with the shorter principal axis of
the DM halo. It is also assumed a triaxiality ratio of the DM halo,
$\tau=(a^2-b^2)/(a^2-c^2)=0.84$, where $a>b>c$, which corresponds to
an average radius of about 1 Mpc. Under these conditions, as found out
by \citet{MC03}, the morphology of the resulting galaxy is highly
related to the initial properties of the halo, which drive the galaxy
formation and evolution. In particular, the dynamical timescales of
stars and gas are quite different and, as a consequence, the two
components become decoupled during the evolution of the
system.\\ Here, we present the synthetic SED which accounts for
chemical evolution, stellar emission, internal extinction and
re-emission by dust in a self-consistent way, as described in
\citep{Maz92, Maz95}. This allows us to extend the SED over four orders
of magnitude in wavelength, i.e., from 0.1 to 1000 $\mu$m.  So the
model self-consistently provides morphological, dynamic and
chemo-photometric evolution. Fig. \ref{sed} compares the predicted SED
with the available data for NGC1947. Such a model suggests that star
formation started about 14.5 Gyrs ago, and that the maximum was reached
about 9 Gyrs ago. The average age for the inner stellar generations is
dated at about 3 Gyrs: this could be due to the still efficient inflow
of the gas and is consistent with the findings by
\citet{Ser07}, which included such a galaxy in the sample of
centrally-rejuvenated objects.\\ Dynamical predictions of such SPH
simulations agree with kinematical features observed for NGC1947 (see
Sec. \ref{kinem}): in particular, the model predicts a maximum rotation
velocity of about 20 km/s for the stars and of 120 km/s for the gas,
and a velocity dispersion in the central regions of about 130 km/s.

- \emph{Conclusions} -

Putting all the above evidences together, we now try to address which
are the observational aspects that can help to disentangle in a non
ambiguous way the possible formation scenarios for NGC1947 discussed
above.

We have found that  an unequal-mass merging of two disk galaxies,
with a mass ratio in the range 3:1 - 4:1, could account for the
average structure and kinematics observed for NGC1947. But this
scenario fails to reconcile $\it{i})$ the epoch of the merging
event with the estimates of stellar population ages in NGC1947;
and $\it{ii})$ the gas kinematics. According to the simulations
(\citealt{Bour05}), after 3 Gyrs from the merging event, the outer
regions of the remnant are characterized by typical signs of the
past interaction, like tidal tails and/or shells.  Such features
are not observed in NGC1947.  To account for this, the merger
would have occurred very long ago ($\sim 10$ Gyrs) and the ring
should be very long-lived: these predictions are not consistent
with both $\it{i})$ the upper limit to the epoch at which the
interaction happened for NGC1947, i.e.  $\sim 3$ Gyrs (see
Sec.\ref{prof}) and $\it{ii})$ the warped structure of the ring,
which suggests that the polar component has not reached a stable
configuration yet (see sec.\ref{kinem}). Therefore, we can
reasonably rule out the merging scenario for the formation of
NGC1947 and we can regard the recent polar accretion from a gas
rich donor galaxy as one of the most probable interaction event in
the formation history of this galaxy. \\ We have also discussed
how such a peculiar object may be the result of processes of
galaxy formation and evolution under particular initial
conditions, linked to the ratio between baryonic and dark matter:
the SPH simulations by \citet{MC03} can reasonably account for the
observed photometry and kinematics of the galaxy. We aim to
further investigate in detail such aspect by comparing the
predictions of such a model and observations, but they are beyond
the aims of the present work and they will be the subject of a
forthcoming paper.


\section*{Acknowledgments}
The authors wish to thank the anonymous referee whose comments
and suggestions greatly improved the presentation of this work.  The
authors are very grateful to F. Bournaud for many useful discussions
and suggestions. E.I. wish to thank E. Pompei for the support given
during the data acquisition. This work is based on observations made
with ESO Telescopes at the Paranal Observatories under programme ID
$<70.B-0253(A)>$ and $<74.B-0626(A)>$.


\bibliography{N1947_bib_mnras}

\bsp

\label{lastpage}

\end{document}